\documentclass[preprint2]{proto_jb}
\usepackage{times}

\newcommand{\BP}{Ballesteros-Paredes}
\newcommand{\cs}{c_{\rm s}}

\newcommand{\kms}{{\rm km~s}^{-1}}
\newcommand{\lj}{L_{\rm J}}
\newcommand{\ls}{\lambda_{\rm s}}

\newcommand{\ns}{n_{\rm s}}
\newcommand{\nsl}{n_{\rm sl}}

\newcommand{\pcc}{{\rm cm}^{-3}}
\newcommand{\psc}{{\rm cm}^{-2}}

\newcommand{\tff}{t_{\rm ff}}
\newcommand{\VS}{V{\' a}zquez-Semadeni}

\def\'#1{\ifx#1i{\accent"13\i}\else{\accent"13#1}\fi}

\def\alamenos#1{$^{-#1}$}

\begin{document}

\title{\textbf{\LARGE Molecular Cloud Turbulence and Star Formation}}

\author{\textbf{\large Javier \BP}}
\affil{\small\em Universidad Nacional Aut\'onoma de M\'exico}
\author{\textbf{\large Ralf S. Klessen}}
\affil{\small\em  Astrophysikalisches Institut Potsdam}
\author{\textbf{\large Mordecai-Mark Mac Low}}
\affil{\small\em American Museum of Natural History at New York}
\author{\textbf{\large Enrique \VS}}
\affil{\small\em Universidad Nacional Aut\'onoma de M\'exico}

\begin{abstract}
\baselineskip = 11pt
\leftskip = 0.65in 
\rightskip = 0.65in
\parindent=1pc

   {\small We review the properties of turbulent molecular clouds (MCs),
   focusing on the physical processes that influence star formation (SF).
MC formation appears to occur during large-scale compression of the 
diffuse ISM driven by supernovae, magnetorotational 
instability, or gravitational instability  in
galactic disks of stars and gas.
   The compressions generate turbulence
that can accelerate
   molecule production and produce the
   observed morphology. We then review the properties of MC turbulence,
   including density enhancements
observed as clumps and cores,
   magnetic field structure,
   driving scales, the relation to observed scaling relations, and the
   interaction with gas thermodynamics.  We argue that MC cores are
   dynamical, not quasistatic, objects with relatively short lifetimes
   not exceeding a few megayears.  We review their morphology, magnetic
   fields, density and velocity profiles, and virial budget.  Next, we
   discuss how MC turbulence controls SF.  On global
   scales turbulence prevents monolithic collapse of the clouds; on small
   scales it promotes local collapse. We discuss its effects on the SF
   efficiency, and critically examine the possible relation between the
   clump mass distribution and the initial mass function, and then turn
   to the redistribution of angular momentum during collapse and how
   it determines the multiplicity of stellar systems. Finally, we discuss
   the importance of dynamical interactions between protostars in dense
   clusters, and the effect of the ionization and winds from those
   protostars on the surrounding cloud.
We conclude that
   the interaction of self-gravity and turbulence controls MC formation
   and behavior, as well as the core and star formation processes within
   them.}

\mbox{}\vspace{0.5cm}

\end{abstract}  

\section{\textbf{INTRODUCTION}}

Star formation occurs within molecular clouds (MCs). These exhibit
supersonic linewidths, which are interpreted as evidence for
supersonic turbulence \citep{ZUC74b}.  Early studies considered this
property mainly as a mechanism of MC support against gravity.  In more
recent years, however, it has been realized that turbulence is a
fundamental ingredient
of MCs, determining properties such as their morphology, 
lifetimes, rate of star formation, etc.

Turbulence is a multiscale phenomenon in which kinetic energy cascades
from large scales to small scales. The bulk of the specific kinetic energy
remains at large scales.  
Turbulence appears to be
dynamically important from scales of whole MCs down to cores
\citep[e.g., ][]{LAR81, BAL99b,MAC04}. Thus, early microturbulent
descriptions postulating that turbulence {\it only} acts on small scales
did not capture major effects at large scales such as cloud and core
formation by the turbulence. 

Turbulence in the warm diffuse interstellar medium (ISM) is transonic,
with both the sound speed and the non-thermal motions being $\sim
10~\kms$ \citep{KH87,HT03}, while within MCs it is highly supersonic,
with Mach numbers ${\cal M} \approx 5$--20
\citep{ZUC74}.  Both media are highly compressible. 
Hypersonic velocity fluctuations in the roughly
isothermal molecular gas produce large density enhancements at shocks. The
velocity fluctuations in the warm diffuse medium, despite being only
transonic, can 
still
drive large density enhancements because they can
push the medium into a thermally unstable regime in which the gas
cools rapidly into a cold, dense regime \citep{HP99}. In general, the
atomic gas 
responds close to isobarically to dynamic compressions for densities
$0.5 \lesssim n/\pcc \lesssim 40$ \citep{GAZ05}.
We argue in this review that MCs form from dynamically evolving,
high-density features in the diffuse ISM. Similarly,
their internal substructure of clumps and cores are also transient
density enhancements continually 
changing their shape and even the material contained in the turbulent
flow, behaving as something between discrete objects and waves
\citep{VAZ96}. Because of their higher density, some clumps and cores
become gravitationally unstable and collapse to form stars.

We here discuss the main advances in our understanding of the turbulent
properties of MCs and their implications for star formation since the
reviews by \citet{VAZ00} and \cite{MAC04},  
proceeding from large (giant MCs) to small (core) scales.

\section{\textbf{MOLECULAR CLOUD FORMATION}}  \label{sec:mc_form}

The questions of how MCs form and what determines their physical
properties have remained unanswered until recently \citep[e.g.,][]
{ELM91b, 
BLI99}. Giant molecular clouds (GMCs) have gravitational
energy far exceeding their thermal energy \citep{ZUC74}, although
comparable to their turbulent \citep{LAR81} and magnetic energies
\citep{MYE88a,CRU99,BOU01,CRU04}.  This near equipartition of energies
has traditionally been interpreted as indicative of approximate virial
equilibrium, and thus of general stability and longevity of the clouds
\citep[e.g.,][]{MCK93,BLI99}.  In this picture, the fact that MCs have
thermal pressures exceeding that of the general ISM by roughly one
order of magnitude \citep[e.g.,][]{BLI99} was interpreted as a
consequence of their being strongly self-gravitating
\citep[e.g.,][]{McK95}, while the magnetic and turbulent energies were
interpreted as support against gravity.  Because of the interpretation
that their overpressures were due to self-gravity, MCs could not be
incorporated into global ISM models based on thermal pressure
equilibrium, such as those by \cite{FIE69, MCK77, WOL95}.

Recent work suggests instead that MCs are likely to be
{\em transient}, dynamically evolving features produced by compressive
motions of either gravitational or turbulent origin, or some combination
thereof. In what follows we first discuss these two
formation mechanisms, and then discuss how they can give rise
to the observed physical and chemical properties of MCs.

\subsection{\textbf{Formation mechanisms}} \label{sec:MC_form_mech}




Large-scale gravitational instability in the combined medium of the
collisionless stars 
and the collisional gas
appears likely to be the main driver of GMC formation in
galaxies. In the Milky Way, MCs, and particularly GMCs, are observed
to be concentrated towards the spiral arms
\citep{
LSKM01,BR05,STA05}, which are the first manifestation of gravitational
instability in the combined medium.  We refer to the instability
parameter for stars and gas combined \citep{RAF01} as the Toomre
parameter for stars and gas $Q_{sg}$. As $Q_{sg}$ drops below unity,
gravitational instability
drives spiral density waves \citep{LIN64,LIN69}.
High-resolution numerical simulations of the process then show the
appearance of regions of local gravitational collapse in the arms,
with the timescale for collapse depending exponentially on the value
of $Q_{sg}$
\citep{LI05b}. In more strongly gravitationally unstable regions
towards the centers of galaxies, collapse occurs more generally, with
molecular gas dominating over atomic gas, as observed by
\citet{WON02}.

The multi-kiloparsec scale of spiral arms driven by gravitational
instability suggests that this mechanism should preferentially form
GMCs, consistent with the fact that GMCs are almost exclusively
confined to spiral arms in our galaxy \citep{STA05}. Also, stronger
gravitational instability causes collapse to occur closer to the disk
midplane, producing a smaller scale height for GMCs \citep{LI05b}, as
observed by \citet{DYB04} and \citet{STA05}.  In gas-poor galaxies
like our own, the self-gravity of the gas in the unperturbed state is
negligible with respect to that of the stars, so this mechanism
reduces to the standard scenario of the gas falling into the potential
well of the stellar spiral and shocking there \citep{ROB69}, with its
own self-gravity only becoming important as the gas is shocked and
cooled (Section~\ref{sec:orig_MC_props}).

A second mechanism of MC formation operating at somewhat smaller scales
(tens to hundreds of pc) is the ram pressure from supersonic flows,
which can be produced by a number of different sources.  
%
%
Supernova explosions drive blast waves and
superbubbles into the ISM. Compression and gravitational collapse can
occur in discrete superbubble shells \citep[][]{MCC87}.
The ensemble of supernova remnants, superbubbles and expanding HII
regions in the ISM drives a turbulent flow \citep{VAZ95,PAS95,ROS95,
KOR99, GAZ99, AVI00, AVI01, SLY05, MAC05, DIB05, JOU06}.  Finally, the
magnetorotational 
instability \citep{BAL98} in galaxies can drive
turbulence with velocity dispersion close to that of supernova-driven
turbulence even in regions far from recent star formation
\citep[][]{SEL99,KIM03, DZI04, PIO05}. Both of these sources of turbulence
drive flows with velocity 
dispersion $\sim 10~\kms$, similar to that observed
\citep[][]{KH87, DIC90, HT03}.  The high-$\cal M$ tail of the shock
distribution in the warm medium involved in this turbulent flow can
drive MC formation.

Small-scale, scattered, ram-pressure compressions in a globally
gravitationally stable medium can produce clouds that are
globally supported against collapse, but that still undergo local
collapse in their densest substructures.  These events involve
a small fraction of the total cloud mass and are characterized by a
shorter free-fall time than that of the parent cloud because of the
enhanced local density \citep[][]{SAS73, ELM93, PAD95, VAZ96, BAL99a,
  BAL99b, KLE00b, HEI01, VAZ03, VAZ05,  LI04b, NAK05, CLA05}.
Large-scale gravitational contraction, on the other hand, can produce
strongly bound GMCs in which gravitational collapse proceeds
efficiently.  
In either case, mass that does not collapse sees its density reduced,
and may possibly remain unbound throughout the evolution
\citep{BON06}. The duration of the entire gas accumulation process to
form a GMC by either kind of compressions may be $\sim$ 10--20 Myr,
with the duration of the mainly {\em molecular} phase probably
constituting only the last 3--5 Myr \citep[][]{BAL99a, HAR01, HAR01b, HAR03}.

Thus, GMCs and
smaller MCs may represent two distinct populations: One formed by the
large-scale gravitational instability in the spiral arms, and
concentrated toward the 
midplane because of the extra gravity in the arms; the other, formed
by more scattered turbulent compression events, probably driven by supernovae,
and distributed similarly to the turbulent atomic gas.

\subsection{\textbf{Origin of Molecular Cloud Properties}}
\label{sec:orig_MC_props}

Large-scale compressions can account for the physical and chemical
conditions characteristic of MCs, 
whether the compression comes from gravity or ram pressure.
\citet[]{HAR01b} estimated that
the column density thresholds for becoming self-gravitating,
molecular, and magnetically supercritical are very similar \citep[see
also][] {FRA86}, $N \sim 1.5 \times 10^{21} (P_{\rm e}/k)_4^{1/2}
\psc$, where $(P_{\rm e}/k)_4$ is the pressure external to the cloud
in units of $10^4$ K $\pcc$.  

MCs are observed to be magnetically critical \citep[][]{CRU03}, yet the
diffuse atomic ISM is
reported to be 
subcritical \citep[e.g.,][]{CRU03, HT05},
apparently violating mass-to-magnetic flux conservation. 
However, the formation of a GMC of $10^6 M_\sun$ out of diffuse gas at
1 $\pcc$ requires accumulation lengths $\sim 400$ pc, a scale over
which the diffuse ISM is magnetically critical \citep{HAR01b}.  Thus,
reports of magnetic subcriticality need to be accompanied by an
estimate of the applicable length scale.

\begin{figure}[h] 
 \epsscale{0.8}
 \plotone{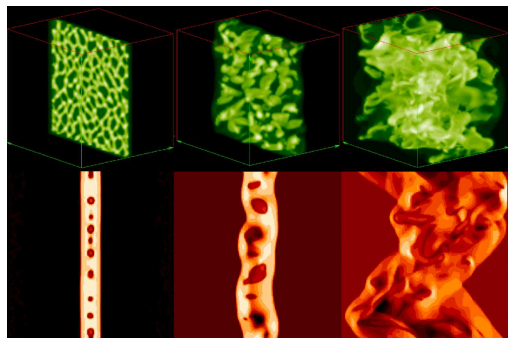}
 \caption{\small \emph{Top:} Projections of the density field in a
 simulation of MC formation by the collision of convergent streams,
 each at a Mach number ${\cal M} = 2.4$. From left to right, 
 times $t=2$, 4 and 7.33 Myr are shown, illustrating how the
 collision first forms a thin sheet that then fragments, becomes
 turbulent, and thickens, until it becomes a fully three-dimensional
 cloud. \emph{Bottom:} Cross sections through the pressure field at the
 $xy$ plane at half the $z$ extension of the simulation. The color
 code ranges from $P/k =525$ (black) to 7000 K
 $\pcc$ (white). \citep[From][]{VAZ06}.
\label{fig:collision} }

\end{figure}

Large-scale compressions also appear capable of producing the observed
internal turbulence in MCs.  \citet{VIS94} showed analytically that
bending modes in isothermal shock-bounded layers are {\em nonlinearly}
unstable. This analysis has been confirmed numerically \citep{HUN86,
STE92, BLO96, WAL98, WAL00},
demonstrating that shock-bounded, radiatively cooled layers can become
unstable and develop turbulence.  Detailed models of the compression
of the diffuse ISM show that the compressed post-shock gas undergoes
both thermal and dynamical instability, fragmenting into cold, dense
clumps with clump-to-clump velocity dispersions of a few kilometers
per second, supersonic with respect to their internal sound speeds
\citep{KOY02,INU04,AUD05, HEI05, VAZ06}.  
\citet{INU04} included magnetic fields, with similar results.
\citet{VAZ06} further showed that 
gas with MC-like densities occurs in regions overpressured by dynamic
compression to factors of as much as 5 above the mean thermal pressure
(see Fig.~\ref{fig:collision}).
These regions are the density peaks in a turbulent flow, rather than
ballistic objects in a diffuse substrate.  Very mildly supersonic
compressions can form thin, cold atomic sheets like those observed by
\citet{HT03} rather than turbulent, thick objects like MCs. Depending on
the inflow Mach number, turbulence can take 5--100~Myr to develop.

A crucial question is whether molecules can form on the short
timescales (3--5 Myr) implied by the ages of the stellar populations
in nearby star-forming regions \citep[][]{BAL99b,HAR01,HAR01b}. H$_2$ molecule
formation on grains occurs on timescales $t_{\rm f} = 10^9\mbox{
yr}/n$ \citep{HS71,JUR75}.
\cite{PAV02} showed that, in a turbulent flow, H$_2$ formation
proceeds fastest in the highest density enhancements, yet the bulk of
the mass passes through those enhancements quickly.  
\citet{GLO05} confirm that this mechanism produces widespread
molecular gas at average densities of order $10^2$~cm$^{-3}$ in
compressed, turbulent regions. This may explain recent observations of
diffuse ($n_{\rm H} \leq$ 30~cm\alamenos 3) cirrus clouds that exibit
significant fractions of H$_2$ \citep[$\sim$ 1\%--30\%][]{GIL05}. Rapid
H$_2$ formation suggests that a more relevant MC  formation timescale is 
that required to accumulate a sufficient column density for extinction
to allow CO formation.  Starting from H~I with $n\sim 1$ cm$^{-3}$,
and $\delta v\sim 10$ km/sec, this is $\sim$ 10--20 Myr \citep{BER04}.


In the scenario we have described, MCs are generally not in
equilibrium, but rather evolving secularly. They start as atomic gas
that is compressed, increasing its mean density. The atomic gas may or
may not be initially self-gravitating,
but in either case the compression causes the gas to cool
via thermal instability and to develop turbulence via dynamical
instabilities. Thermal instability and turbulence promote
fragmentation and, even though the self-gravity of the cloud as a
whole is increased due to cooling and the compression, the free-fall
time in the fragments is shorter, so collapse proceeds locally,
preventing monolithic collapse of the cloud and thus reducing the star
formation efficiency (Section~\ref{sec:SFE}). The near equipartition
between their gravitational, magnetic and turbulent kinetic energies
is not necessarily a condition of equilibrium \citep{MAL90, VAZ95,
BAL97, BAL99b}, and instead may simply indicate that 
self-gravity has become important either due to large-scale
gravitational instability or local cooling.

\section{\textbf{PROPERTIES OF MOLECULAR CLOUD TURBULENCE}}\label{sec:MC_turb}


In a turbulent flow with velocity power spectrum having negative
slope, the typical velocity difference between two points decreases as
their separation decreases.  Current observations give $\Delta v
\approx 1 ~\kms \left(R/[1~{\rm pc}]\right)^{\beta}$ with $\beta
\simeq$ 0.38--0.5 \citep[e.g., ][]{LAR81,BLI93,HEY04}.  This scaling
law implies that the typical velocity difference across separations
$\ell > \ls \sim 0.05$ pc is supersonic, while it is subsonic for
$\ell < \ls$, where we define $\ls$ as the sonic scale. Note that
dropping to subsonic velocity is unrelated to reaching the dissipation
scale of the turbulence, though there may be a change in slope of the
velocity power spectrum at $\ls$.

The roughly isothermal supersonic turbulence in MCs at scales $\ell >
\ls$ produces density enhancements that appear to constitute the
clumps and cores of MCs \citep{WEI51,
SAS73,ELM93,PAD95,BAL99b}, since the
density jump associated with isothermal shocks is $\sim {\cal
M}^2$. Conversely, the subsonic turbulence at $\ell < \ls$ within
cores does not drive further fragmentation, 
and
is less important than thermal pressure in resisting self-gravity
\citep{PAD95,VAZ03}. 


\subsection{\textbf{Density Fluctuations and Magnetic Fields}}
\label{sub:den_mag}

The typical amplitudes and sizes
of density fluctuations driven by supersonic turbulence determine
if they will become gravitationally unstable and collapse to form
one or more stars. One of the simplest statistical indicators
is the density probability distribution function (PDF). For nearly
isothermal flows subject to random accelerations, the density PDF
develops a log-normal shape \citep{VAZ94,PAD97b,PAS98,KLE00c}.
This is the expected shape if the density
fluctuations are built by successive passages of shock waves at any
given location and the jump amplitude is independent of the local
density \citep{PAS98,NOR99}. Numerical studies in
three dimensions suggest that the standard deviation of the logarithm of
the density fluctuations $\sigma_{\log \rho}$ scales with the
logarithm of the Mach number \citep{PAD97b,MAC05}. 

Similar features appear to persist in the isothermal, ideal MHD case
\citep{OST99,OST01,FAL02}, with the presence of the field not
significantly affecting the shape of the density PDF except for particular
angles between the field and the direction of the MHD wave propagation
\citep{PAS03}.  
In general, $\sigma_{\log \rho}$ tends to decrease with increasing
field strength $B$, although it exhibits slight deviations from
monotonicity.  Some studies find that intermediate values of $B$
inhibit the production of large density fluctuations better than
larger values \citep{PAS95,OST01,BAL02b,VAZ05c}.
This may be due  to a more
isotropic behavior of the magnetic pressure at intermediate values of
$B$ \citep[see also][]{HEI01}.

An important observational \citep{CRU03} and numerical
\citep{PAD99,OST01,PAS03} result is that the magnetic field strength
appears to be uncorrelated with the density at low-to-moderate
densities or strong fields, while a correlation appears at high
densities or weak fields. \citet{PAS03} suggest this occurs because
the slow mode dominates for weak fields, while the fast mode dominates for
strong fields.

Comparison of observations to numerical models has led \citet{PAD99},
\citet{PAD03} and \citet{PAD04a} to strongly argue that MC turbulence
is not just supersonic but also super-Alfv\'enic with respect to the
initial mean field.  \citet{PAD99} notes that such turbulence can
easily produce trans- or even sub-Alfv\'enic cores. Also, at advanced
stages, the total magnetic energy including fluctuations in driven
turbulence approaches the equipartition value \citep{PAD03}, due to
turbulent amplification of the magnetic energy \citep{OST99,SCH02}.
Observations, on the other hand, are generally interpreted as  
showing that MC cores are magnetically critical or moderately
supercritical and therefore trans- or super-Alfv\'enic if the
turbulence is in equipartition with self-gravity
(\citealt{BER92,CRU99,BOU01,CRU03}; see also the chapter by {\em di
Francesco et al.}).  However, the observational
determination of the relative importance of the magnetic field
strength in MCs depends on their assumed geometry, and the clouds
could even be strongly super-critical \citep{BOU01}. Unfortunately,
the lack of self-gravity, which could substitute for stronger
turbulence in the production of denser cores in the simulations by
Padoan and coworkers, implies that the evidence in favor of
super-Alfv\'enic MCs remains inconclusive.

\subsection{\textbf{Driving and Decay of the Turbulence}}
\label{sub:turb-props} 

Supersonic turbulence should decay in a crossing time of the driving
scale \citep{GOL74b,FIE78}.  For a number of years it was thought that
magnetic fields might modify this result \citep{ARO75}, but it has
been numerically confirmed that both hydrodynamical and MHD turbulence
decay in less than a free-fall time, whether or not an isothermal
equation of state is assumed \citep{STO98,MAC98,PAD99,BIS00, AVI01, PAV02}.
%
Decay does
proceed more slowly if there are order of magnitude imbalances in
motions in opposite directions \citep{MAR01,CHO02},
though this seems unlikely to occur in MCs.  


Traditionally, MCs have been thought to have lifetimes much larger
than their free-fall times \citep[e.g.,][]{BLI80}. However, recent work
suggests that MC lifetimes may be actually comparable or shorter than
their free-fall times \citep{BAL99a, HAR01b, HAR03}, and that star
formation occurs within a single crossing time at all scales
\citep{ELM00b}. If true, and if clouds are destroyed promptly after
star formation has occurred \citep{FUK99,YAM01}, this
suggests that turbulence is ubiquitously observed simply because it is
produced together with the cloud itself (see\ Section~\ref{sec:orig_MC_props}) and it does not have time to decay
afterwards.


Observations of nearby MCs show them to be self-similar up to the largest
scales traced by molecules \citep{MAC00,OSS01,OSS02,BRU03,HEY04},
suggesting that they are driven from even larger scales (see also Section~\ref{dens_vel_fields:sec}). If the
turbulent or gravitational compressions that form MCs drive the
turbulence, then the driving scale would indeed be larger than the
clouds.

\subsection{Spectra and Scaling Relations}\label{sub:spectra_scal}

Scaling relations between the mean density and size, and between
velocity dispersion and size \citep{LAR81} have been discussed in
several reviews \citep[e.g., ][]{VAZ00, MAC04, ELM04}.  In order to
avoid repetition, we comment in detail only on the caveats not discussed there.

The Larson relation $\langle\rho\rangle \sim R^{-1}$ 
implies a constant column density.  However,
observations tend to have a limited dynamic range, and thus to select
clouds with similar column densities \citep{KEG89, SCA90, SCH04b}, as
already noted by Larson himself.  Numerical simulations of both the
diffuse ISM \citep{VAZ97} and of MCs \citep{BAL02b} suggest that
clouds of lower column density exist.  However, when the observational
procedure is simulated, then the mean density-size relation appears,
showing that it is likely to be an observational artifact.

\citet{ELM04} suggested that the mass-size relation found in the
numerical simulations of \citet{OST01} supports the reality of the
$\langle\rho \rangle$-$R$ relation.  However, those authors defined
their clumps using a procedure inspired by observations, focusing on
regions with higher-than-average column density.  As in noise-limited
observations, this introduces an artificial cutoff in column density,
which again produces the appearance of a $\langle \rho \rangle$-$R$
relation.

Nevertheless, the fact that 
the column density threshold for a cloud to become molecular is
similar to that for becoming self-gravitating \citep{FRA86, HAR01b} may
truly imply a limited column density range for \emph{molecular} gas,
since clouds of lower column density are not molecular, while clouds
with higher column densities collapse quickly. MCs constitute only the
tip of the iceberg of the neutral gas distribution.

The $\Delta v$-$R$ relation is better established.
However, it does not appear to depend on self-gravitation as has often
been argued.  Rather
it appears to be a measurement of the second-order structure function
of the turbulence \citep{ELM04}, which has a close connection with the energy spectrum 
$E(k) \propto k^n$.  The relation between the spectral index $n$ and
the exponent in the velocity dispersion-size relation 
$\Delta v \propto R^{\beta}$,
is $\beta =
-(n+1)/2$.  Observationally, the value originally found by
\citet{LAR81}, $\beta \approx 0.38$, is close to the value expected
for incompressible, Kolmogorov turbulence ($\beta = 1/3$,
$n=-5/3$). More recent surveys of whole GMCs tend to give values
$\beta \sim$ 0.5--0.65 \citep{BLI93, MIZ01, HEY04}, with smaller
values arising in low surface brightness regions of clouds
\citep[][$\beta \sim 0.4$]{FAL92} and massive core surveys
\citep[][$\beta \sim$ 0--0.2, and very poor
correlations]{CAS95,PLU97}. The 
latter may be affected by intermittency effects, stellar feedback, or
strong gravitational contraction. Numerically, simulations of MCs and
the diffuse ISM \citep{VAZ97, OST01, BAL02b} generally exhibit very
noisy $\Delta v$-$R$ relations both in physical and observational
projected space, with slopes $\beta \sim$ 0.2--0.4. The scatter is
generally larger at small clump sizes, both in physical space, perhaps
because of intermittency, and in projection, probably because of
feature superposition along the line of sight.

There is currently no consensus on the
spectral index for compressible flows. A value
$n=-2$ ($\beta = 1/2$) is expected for shock-dominated flows
\citep[e.g., ][]{ELS76}, while 
\citet{BOL02a} has presented a theoretical model
suggesting that $n \approx -1.74$, and thus $\beta \approx
0.37$.  The predictions of this model for the higher-order 
structure function scalings appear to be confirmed numerically 
\citep{BOL02b,PAD04b, JOU06, AVI06}. 
However, the model assumes incompressible Kolmogorov scaling
at large scales on the basis of simulations driven with purely
solenoidal motions \citep{BOL02b}, which does not appear to agree with
the behavior seen by ISM simulations \citep[e.g., ][]{AVI00, JOU06}.

Numerically, 
\citet{PAS95} reported $n = -2$ in their two-dimensional simulations
of magnetized, self-gravitating turbulence in the diffuse ISM. More
recently, \citet{BOL02b} have reported $n=-1.74$ in isothermal,
non-self-gravitating MHD simulations, while other studies find that
$n$ appears to depend on the rms Mach number of the flow in both
magnetic and non-magnetic cases \citep{CHO03,BAL06}, approaching $n
= -2$ at high Mach numbers.

\subsection{\textbf{Thermodynamic Properties of the Star-Forming Gas}}
\label{sub:EOS}


While the atomic gas is
tenuous and warm, with densities $1 < n < 100\, \pcc$
and temperatures 
$100\,$K$ < T < 5000\,$K,  
MCs have $10^2 < n < 10^3\,$cm$^{-3}$, and locally $n >
10^6\,$cm$^{-3}$ in dense cores. The kinetic temperature inferred from
molecular line ratios is typically about $10\,$K for dark, quiescent
clouds and dense cores, but can reach 50--100~K in regions heated by
UV radiation from high-mass stars \citep{KUR00}.

MC2 temperatures are determined by the balance between heating and
cooling, which in turn both depend on the chemistry and dust content.  
Early studies predicted that the equilibrium
temperatures in MC cores should be 10--20~K,
tending to be lower at the higher densities
\citep[e.g.,][]
{HAY66,LAR69,LAR73,GOL78}.  Observations
generally agree with these values \citep{JIJ99},
prompting most theoretical and numerical star-formation studies to
adopt a simple isothermal description, with  $T \sim 10\,$K.

In reality, however, the
temperature varies by factors of two or three above and below the
assumed constant value \citep[][]{GOL01,LAR05}.
This variation can be described by an 
effective equation of state (EOS) derived from detailed
balance between heating and cooling when cooling is fast
\citep{VAZ96,PAS98,SCA98b,SPA00,LI03,SPA05}.  This EOS can often be
approximated by a polytropic law of the form $P \propto \rho^\gamma$,
where $\gamma$ is a parameter.  For molecular clouds, there are three
main regimes for which different polytropic EOS can be defined.  {\em
(a)} For $n \lesssim 2.5\times 10^5\,\mathrm{cm}^{-3}$, MCs are
externally heated by cosmic rays or photoelectric heating, and cooled
mainly through the emission of molecular (e.g., CO, H$_2$O) or atomic
(C II, O , etc.) lines, depending on ionization level and chemical
composition \citep[e.g., ][]{GEN91}. 
The strong dependence of the cooling rate on density yields an
equilibrium temperature that decreases with increasing density.  In
this regime, the EOS can be approximated by a polytrope with $\gamma =
0.75$ \citep{KOY00}.  Small, dense cores indeed are observed to have
$T \sim 8.5\,$K at a density $n \sim 10^{5}\,\mathrm{cm}^{-3}$
\citep{EVA99}.
{\em (b)} At $n > 10^6\,$cm$^{-3}$, the gas becomes
thermally coupled to dust grains, which then control the temperature
by their far-infrared thermal emission.  In this regime, the effective
polytropic index is found to be $1 < \gamma < 1.075$.
\citep{SCA98b,SPA00,LAR05}. 
The temperature is predicted to reach a minimum
of $5\,$K \citep{LAR05}.
Such cold gas is difficult to observe, but observations have been
interpreted to suggest central temperatures between $6\,$K and
$10\,$K for the densest observed prestellar cores.
{\em (c)} Finally, at $n > 10^{11}\,
\mathrm{cm}^{-3}$, the dust opacity increases, 
causing the temperature to rise rapidly.  This results in an opacity
limit to fragmentation that is somewhat below $0.01\,M_{\odot}$
\citep{LOW76, MAS00}. The transition from $\gamma < 1$ to $\gamma
\gtrsim 1$ in molecular gas may determine the characteristic stellar mass
resulting from gravoturbulent fragentation (Section~\ref{sub:IMF}).


\section{\textbf{PROPERTIES OF MOLECULAR CLOUD CORES}}\label{sec:MC_cores}

\subsection{\textbf{Dynamical Evolution}}\label{sec:dyn_evol}

\noindent{\em 4.1.1. Dynamic or Hydrostatic Cores? \label{sec:dyn_stat}}
%
The first successful models of MCs and their cores used hydrostatic
equilibrium configurations as the starting point (e.g.,
\citealt{DEJ80, SHU87, MOU91a}; see also Section~5.1). However,
supersonic turbulence generates the initial density enhancements from
which cores develop
\citep[e.g.,][]{SAS73, PAS88, VAZ94, PAD97, PAS98, PAD99, KLE00b,
HEI01, PAD01a, FAL02, 
PAS03, BAL03, HAR03b, KLE05, KIM05, BER05}.  Therefore, they do not
necessarily approach hydrostatic equilibrium at any point in
their evolution \citep{BAL99b}. For
the dynamically-formed cores to settle into equilibrium
configurations, it is necessary that the equilibrium be {\em stable}.
In this subsection we discuss the conditions necessary for the
production of stable self-gravitating equilibria, and whether they are
likely to be satisfied in MCs.

Known hydrostatic stable equilibria in non-magnetic, self-gravitating
media require confinement by an external pressure to prevent expansion
\citep{BER92, HAR98, HAR01b, VAZ05}.  The most common assumption is a
discontinuous phase transition to a warm, tenuous phase that provides
pressure without adding weight (i.e., mass) beyond the core's
boundary. The requirement of a confining medium implies that the
medium must be two-phase, as in the model of \cite{FIE69} for the
diffuse ISM. Turbulent pressure is probably not a suitable confining
agent because it is large-scale, transient, and anisotropic, thus
being more likely to cause distortion than confinement of the cores
\citep{BAL99b,KLE05}.

Bonnor-Ebert spheres 
are stable for a ratio of
central to external 
density $< 14$.  Unstable equilibrium solutions exist that dispense
with the confining medium by pushing the boundaries to infinity, with
vanishing density there.  Other, non-spherical, equilibrium solutions
exist \citep[e.g.,][]{CUR00} that do not require an external confining
medium, although their stability remains an open question.  A
sufficiently large pressure increase, such as might occur in a
turbulent flow, can drive a stable, equilibrium object unstable.



For mass-to-flux ratios below a critical value,
the equations of ideal MHD without ambipolar diffusion (AD) predict
unconditionally stable equilibrium solutions, in the sense that no
amount of external compression can cause collapse \citep{SHU87}.
However, these configurations still require a confining thermal or
magnetic pressure, applied at the boundaries parallel to the mean
field, to prevent expansion in the perpendicular direction
\citep{NAK98, HAR01b}.
Treatments avoiding such a confining medium
\citep[e.g., ][]{LIZ89} have used periodic boundary conditions, in
which the pressure from the neighboring boxes effectively confines the
system.
When boundary conditions do not restrict the cores to remain
subcritical, as in globally supercritical simulations, subcritical
cores do not appear \citep{LI04,VAZ05}.

Thus it appears that the boundary conditions of the cores determine
whether they can find stable equilibria. Some specific key questions
are {\em (a)}~whether MC interiors can be considered two-phase media;
{\em (b)}~whether the clouds can really be considered magnetically
subcritical when the masses of their atomic envelopes are taken into
account (cf. Section~\ref{sec:mc_form}), and {\em (c)}~whether the
magnetic field inside the clouds is strong and uniform enough to
prevent expansion of the field lines at the cores. If any of these
questions is answered in the positive, the possibility of nearly
hydrostatic cores cannot be ruled out. 
At this point, they do not appear likely to dominate MCs, though.

\vskip 0.3cm

\noindent{{\em 4.1.2. Core lifetimes. \label{sec:core_lifetimes}}}
%
Core lifetimes are determined using either chemical or statistical
methods. 
Statistical studies use the observed number ratio of starless to
stellar cores in surveys $\nsl/\ns$ (\citealt{BEI86, WAR94, LEE99,
JIJ99, JES00};  see also the chapter by {\em di Francesco et al.}), as a
measure of the time spent by a core in the pre- and protostellar
stages, respectively.  This method gives estimates of
prestellar lifetimes $\tau$ of a few hundred thousand to a few million
years.  Although \cite{JES00} report $6 \leq \tau \leq 17$ Myr, they
likely overestimate the number of starless cores 
because to
find central stars they rely on IRAS measurements of relatively
distant clouds ($\sim$ 350 pc), and so probably miss many of the lower
mass stars.  This is a general problem of the statistical method
\citep{JOR05}.  Indeed, {\em Spitzer} observations have begun to reveal
embedded sources in cores previously thought to be starless
\citep[e.g., ][] {YOU04}.  
Another problem is that if not all present-day starless cores
eventually form stars, then the
ratio $\nsl/\ns$ overestimates the lifetime ratio.
Numerical simulations of isothermal turbulent MCs indeed show that a
significant fraction of the cores (dependent on the exact criterion
used to define them) end up re-expanding rather than
collapsing \citep{VAZ05,NAK05}.

Chemical methods rely on matching the observed chemistry in cores to
evolutionary models \citep[e.g., ][]{TAY96, TAY98, MOR05}.  They
suggest that not all cores will proceed to form stars. 
\citet{LAN00} summarize comparisons of observed molecular abundances
to time-dependent chemical models suggesting typical ages for cores of
$\sim 10^5$~yr. 
Using a similar technique, \citet{JOR05} report a timescale of
$10^{5\pm 0.5}$ yr for the heavy-depletion, dense prestellar stage.

These timescales have been claimed to pose a problem for the model
of AD-mediated, quasi-static, core contraction \citep{LEE99, JIJ99}, as
they amount to only a few local free-fall times ($\tff \equiv L_{\rm
J}/\cs$, $\sim 0.5$ Myr for typical
cores with $\langle n \rangle \sim 5 \times 10^4 \pcc$, where $L_{\rm J}$ is
the local Jeans length, and $\cs$ is the sound speed), rather than the 10--20
$\tff$ predicted by the theory of AD-controlled star formation 
under the assumption of low initial values ($\sim 
0.25$) of the mass-to-magnetic flux ratio (in units of the
critical value) $\mu_0$ \citep[e.g., ][]{CIO95}.  
%
However, \cite{CIO01} have pointed out that the lifetimes
of the ammonia-observable cores predicted by the AD theory are much
lower, reaching values as small as $\sim 2 \tff$ when $\mu_0 \sim
0.8$.  Moreover, nonlinear
effects can enhance the AD rates over the linear estimates
\citep{FAT02, HEI04}.
%
%

%


The lifetimes and evolution of dense cores have recently been investigated
numerically by two groups. \citet{VAZ05}
and \citet{VAZ05c}
considered driven, ideal MHD, magnetically supercritical turbulence with
sustained rms Mach number ${\cal M} \sim 10$, comparable to Taurus,
while \citet{NAK05} considered subcritical, decaying turbulence including
AD, in which the flow spent most of the time at low
$\cal M$, $\sim$ 2--3. Thus, the two setups probably bracket the
conditions in actual MCs. Nevertheless, the lifetimes found in
both studies agree within a factor $\sim 1.5$, ranging from 0.5--10 Myr,
with a median value $\sim$ 2--3 $\tff$ ($\sim 1$--1.5 Myr), 
with the longer timescles corresponding to the decaying cases. 
%
%
Furthermore, a substantial fraction of quiescent, trans- or subsonic
cores, and low star formation efficiencies were found 
(\citealt{KLE05}; see also Section~\ref{dens_vel_fields:sec}),
comparable to those observed in the Taurus molecular cloud (TMC). 
In all cases, the
simulations reported a substantial fraction of re-expanding
(``failed'') cores.

This suggests that the AD timescale, when taking into account the
nearly magnetically-critical nature of the clouds as well as nonlinear
effects and turbulence-accelerated core formation, is comparable to
the dynamical timescale. The notion of quasi-static, long-lived cores
as the general rule is therefore unfounded, except in cases where a
warm confining medium is clearly available \citep[B68 may be an
example; ][]{ALV01}. Both observational and numerical evidence point
towards short typical core lifetimes of a few $\tff$, or a few hundred
thousand to a million years, although the scatter around these values
can be large.

\subsection{\textbf{Core structure}} \label{sec:core_structure}

%


\noindent{{\em 4.2.1. Morphology and magnetic field.}}
%
MC cores in general, and starless cores in particular, have a median
projected aspect ratio $\sim 1.5$ \citep[e.g., ][ and references
therein] {JIJ99}.  It is not clear whether the observed aspect ratios
imply that MC cores are (nearly) oblate or prolate.  \cite{MYE91} and
\cite{RYD96} found that cores are more likely to be prolate than
oblate, while \citet{JON01} found them to be more likely oblate, as
favored by AD-mediated collapse. MHD numerical simulations of
turbulent MCs agree with generally triaxial shapes, with a slight
preference for prolateness \citep{GAM03, LI04}.

The importance of magnetic fields to the morphology and structure of
MCs remains controversial.  Observationally, although some
cores exhibit alignment \citep[e.g., ][]{WOL03}, 
it is common to find misalignments between the polarization angles and
the minor axis of MC cores \citep[e.g., ][]{HEI93, WAR00, CRU04}.  On
the other hand, in modern models of AD-controlled contraction,
triaxial distributions may exist, but the cores are expected to still
be intrinsically flattened along the field \citep{BAS00,JON01, BAS04}.
However, numerical simulations of turbulent magnetized clouds tend to
produce cores with little or no alignment with the magnetic field and
a broad distribution of shapes \citep{BAL02b,GAM03,LI04}. 
 
Finally, a number of theoretical and observational works have studied
the polarized emission of dust from MC cores.  Their main
conclusions are {\it (a)}~cores in magnetized simulations exhibit
degrees of polarization between 1 and 10\%, regardless of whether the
turbulence is sub- or super-Alfv\'enic \citep{PAD01b}; {\em
(b)}~submillimeter polarization maps of quiescent cores do not map the
magnetic fields inside the cores for visual extinctions larger than
$A_V \sim 3$ \citep{PAD01b}, in agreement with observations
\citep[e.g., ][]{WAR00,WOL03, CRU04, BOU04}; {\em (c)}~although the
simplest Chandrasekhar-Fermi method of estimating the mean magnetic
field in a turbulent medium underestimates very weak magnetic fields
by as much as a factor of 150, it can be readily modified to yield
estimates good to a factor of 2.5 for all field strengths
\citep{HEI01b}; and {\em (d)}~limited telescope
resolution leads to a systematic overestimation of field
strength, as well as to an impression of more regular, structureless
magnetic fields \citep{HEI01b}.

\vskip 0.3cm
{\em 4.2.2. Density and Velocity Fields.\label{dens_vel_fields:sec}}
Low-mass MC cores often exhibit column density profiles resembling
those of Bonnor-Ebert spheres (e.g., \citealt{ALV01}; see  see also
the chapter by {\em Lada et al.}) , and trans- or even 
subsonic non-thermal linewidths (e.g., \citealt{JIJ99, TAF04}; see
also the chapter by {\em di Francesco et al.}).  These results
have been traditionally interpreted as evidence that MC cores are in
quasi-static equilibrium, and are therefore long-lived.  This is in
apparent contradiction with the discussion from Section~\ref{sec:dyn_evol}
\citep[see, e.g., the discussion in][]{KET05}, which presented
evidence favoring the view that cores are formed by supersonic
compressions in MCs, and that their lifetimes are of the order of one
dynamical timescale.  It is thus necessary to understand how those
observational properties are arrived at, and whether
dynamically-evolving cores reproduce them.

%
\begin{figure}[h] 
 \epsscale{0.7}
 \plotone{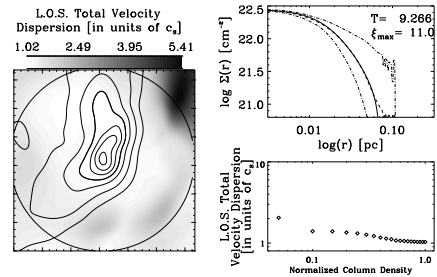}
 \caption{\small Properties of a typical quiescent core formed by
 turbulence in non-magnetic simulations of molecular clouds.
\emph{Left:} Column-density 
 contours on the projected plane of the sky, superimposed on a
 grey-scale image of the {\em total} (thermal plus non-thermal)
 velocity dispersion along each line of sight , showing that the
 velocity dispersion occurs near the periphery of the
 core. \emph{Upper-right:} A fit of a Bonnor-Ebert profile to the
 angle-averaged column density radial profile (\emph{solid line})
 within the circle
 on the map.  The \emph{dash-dotted} lines give the range of variation
 of individual radial profiles before averaging.  \emph{Lower right:}
 Plot of averaged {\it total} velocity dispersion versus column
 density, showing that the core is subsonic.
  (Adapted from \citealt{BAL03} and
\citealt{KLE05}). \label{fig:coherence}}

\end{figure}

\begin{figure}[h] 
 \epsscale{0.545}
 \plotone{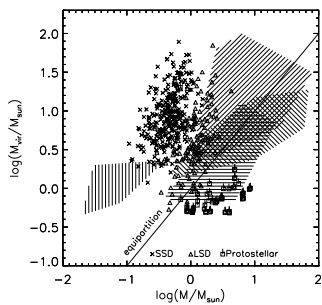}
 \caption{\small \label{fig:coherence2} Virial mass $M_{\rm vir}$
 vs. actual mass $M$ for cores in 
 turbulent models driven at small scales (crosses; ``SSD'') and 
at large scales (triangles; ``LSD'').
 Tailed  squares indicate lower limits on the estimates on $M_{\rm vir}$
 for cores in 
the LSD model containing ``protostellar objects'' (sink particles).
 The solid line denotes the identity $M_{\rm vir} = M$.
Also shown are core survey data by \citet[][ vertical hatching]{MOR05},
\citet[][ horizontal hatching]{ONI02}, \citet[][ $-45^{\circ}$
  hatching]{CAS02}, and \citet[][ $+45^{\circ}$ hatching]{TAC02}. From
\citet{KLE05}.
}

\end{figure}

Radial column density profiles of MC cores are obtained by averaging
the actual three-dimensional structure of the density field 
twice: first, along the line of sight during the observation itself,
and second, during the angle-averaging \citep{BAL03}. These authors
(see also Fig.~\ref{fig:coherence}) further showed that non-magnetic
numerical simulations of turbulent MCs may form rapidly-evolving cores
that nevertheless exhibit Bonnor-Ebert-like angle-averaged column
density profiles, even though they are formed by a supersonic
turbulent compression and are transient structures.  Furthermore,
\cite{KLE05} have shown that $\sim 1/4$ of the cores in their
simulations have subsonic internal velocity dispersions, while $\sim
1/2$ of them are transonic, with velocity dispersions $c_{\rm s} <
\sigma_{\rm turb} \le 2 c_{\rm s}$, where $c_{\rm s}$ is the sound
speed (Fig.~\ref{fig:coherence}). These
fractions are intermediate between those reported by \citet{JIJ99} for
the Taurus and the Perseus MCs. 
These authors also showed that the ratio of virial to actual 
core masses in simulations with large-scale driving
(LSD, 1/2 the simulation size) is reasonably consistent with
observations, while that in simulations with small-scale driving (SSD) is not
(Fig.~\ref{fig:coherence2}), in agreement with the discussion in
Section~\ref{sub:turb-props}. Improved agreement with observations may be
expected for models spanning a larger range of cloud masses.


Quiescent cores in turbulence simulations are not a consequence of
(quasi)-hydrostatic equilibrium, but a natural consequence of the
$\Delta v$-$R$ relation, which implies that the typical
velocity differences across small enough regions ($l_s \lesssim 0.05$ pc)
are subsonic, with $\sigma_{\rm turb} \lesssim c_s$.  Thus, it can be
expected that small cores within turbulent environments may have
trans- or even subsonic non-thermal linewidths.  Moreover, the cores
are assembled by random ram-pressure compressions in the large-scale
flow, and as they are compressed, they trade kinetic compressive
energy for internal energy, so that when the compression is at its
maximum, the velocity is at its minimum \citep{KLE05}. 
Thus,
the observed properties of quiescent low-mass protostellar cores in
MCs do not necessarily imply the existence of strong magnetic fields
providing support against collapse nor of nearly hydrostatic states.

{\em 4.2.3. Energy and Virial Budget.}
The notion that MCs and their cores are in near virial
equilibrium is almost universally encountered in the literature
\citep[e.g., ][ and references therein]{LAR81, MYE88a, MCK99, KRU05}. 
However, what is actually observed is 
energy equipartition between self-gravity and one or more forms of
their internal energy, as \citet{MYE88a}, for example, explicitly
recognize. This by no means implies equilibrium, as the full virial
theorem also contains hard to observe surface and time derivative terms 
\citep{SPI68, PAR79, MCK92}. 
Even the energy estimates suffer from important observational
uncertainties \citep[e.g.,][]{BLI94}. On the other hand, a core driven
into gravitational collapse by an external compression naturally has
an increasing ratio of its gravitational to the kinetic plus magnetic
energies, and passes through equipartition at the verge of collapse,
even if the process is fully dynamic.  This is consistent with the
observational fact that in general, a substantial fraction of clumps
in MCs are gravitationally unbound \citep{MAL90,FAL92,BER92}, a trend
also observed in numerical simulations \citep{VAZ97}.

Indeed, in numerical simulations of the turbulent ISM, the cores
formed as turbulent density enhancements are continually changing
their shapes, and exchanging mass, momentum and energy with their
surroundings, giving large values of the surface and time derivative
terms in the virial theorem (\citealt{BAL95, BAL97, TIL04};  see
\citealt{BAL04} for a review). That the surface terms are large
indicates that there are fluxes of mass, momentum and energy across
cloud boundaries.  Thus, we conclude that observations of
equipartition do not necessarily support the notion of equilibrium,
and that a dynamical picture is consistent with the observations.

\section{\textbf{CONTROL OF STAR FORMATION}}\label{sec:control_SFE}

 \subsection{\textbf{Star Formation Efficiency as Function of
 Turbulence Parameters}}\label{sec:SFE}

The fraction of molecular gas mass converted into stars, called the
star formation efficiency (SFE), is known to be small, ranging from a
few percent for entire MC complexes \citep[e.g., ][]{MYE86} to 10--30\%
for cluster-forming cores \citep[e.g., ][]{LAD03}
and 30\% for starburst galaxies whose gas is primarily molecular
\citep{KEN98}.  \citet{LI06} 
suggest that large-scale gravitational instabilities
in galaxies may directly determine the global SFE in galaxies. They find a
well-defined, nonlinear relationship between the gravitational
instability parameter $Q_{sg}$ (see Section~\ref{sec:MC_form_mech}) and the
SFE.

Stellar feedback reduces the SFE in MCs both by destroying them,
and by driving turbulence within them \citep[e.g., ][]{FRA94, WIL97,
MAT00b, MAT01, MAT02, NAK05}.  Supersonic turbulence has two countervailing
effects that must be accounted for.  
If it has energy comparable to the gravitational energy, it disrupts
monolithic collapse, and if driven can prevent large-scale
collapse. However, these same motions produce strong density
enhancements that can collapse locally, but only involve a small
fraction of the total mass. Its net effect is to
delay collapse compared to the same situation without turbulence
\citep{VAZ99, MAC04}. The magnetic field is only an additional
contribution to the support, rather than the fundamental mechanism
regulating the SFE.

As discussed in Section~\ref{sec:MC_turb}, the existence of a
scaling relation between the velocity dispersion and the size of a cloud
or clump 
implies that within structures sizes smaller than the sonic scale
$\ls$ $\sim 0.05$ pc turbulence is subsonic.  Within structures with
scale $\ell > \ls$, turbulence drives subfragmentation and prevents
monolithic collapse hierarchically (smaller fragments form within
larger ones), while on scales $\ell < \ls$ turbulence ceases to be a
dominant form of support, and cannot drive further subfragmentation
\citep[][]{PAD95, VAZ03}.  (Subsonic turbulence, however, can still
produce weakly nonlinear density fluctuations that can act as seeds
for gravitational fragmentation of the core during its
collapse; Goodwin et al. 2004a) Cores with $\ell < \ls$ collapse if gravity
overwhelms their thermal and magnetic pressures, perhaps helped by ram
pressure acting across their boundaries \citep{HUN82}.

Numerical simulations neglecting magnetic fields showed that the SFE
decreases monotonically with increasing turbulent energy or with
decreasing energy injection scale
\citep[][]{LEO90,KLE00b, CLA04}. A remarkable fact is that the SFE was
larger than zero even for simulations in which the turbulent Jeans
mass \citep[the Jeans mass including the turbulent RMS
velocity; ][]{CHA51a} was larger than the simulation mass, showing that
turbulence can induce local collapse even in globally unbound
clouds \citep{VAZ96, KLE00b, CLA04, CLA05}.

A correlation between the SFE and $\ls$, SFE$(\ls) =
\exp(-\lambda_0/\ls)$, with $\lambda_0 \sim 0.05$ pc, was found in the
numerical simulations of \cite{VAZ03}.  This result provided support
to the idea that the sonic scale helps determine the SFE.
\cite{KRU05b} have further suggested that the other crucial parameter
is the Jeans length $\lj$, and computed, from the probability
distribution of the gas density, the fraction of the gas mass that has
$\lj \lesssim \ls$, which then gives the star formation rate per
free-fall time.  A recent observational study by \citet{HEY05} only
finds a correlation between the sonic scale and the SFE for a subset
of clouds with particularly high SFE, while no correlation is observed
in general. This result might be a consequence of \citet{HEY05} not
taking into account the clouds' Jeans lengths, as required by
\citet{KRU05b}.

\begin{figure}[h] 
 \epsscale{0.75}
 \plotone{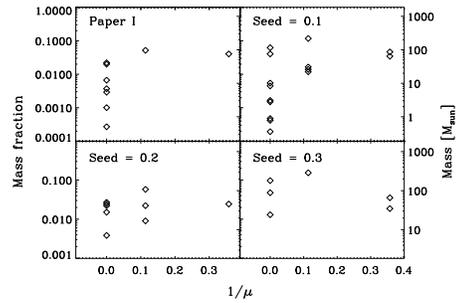}
  \caption{\small \label{fig:magnetic} Masses of the collapsed objects
  formed in four sets of simulations with rms Mach number ${\cal M}
  \sim 10$ and 64 Jeans masses.  Each set contains one non-magnetic
  run and two MHD runs each.  The latter with have initial
  mass-to-flux ration $\mu_0 = 8.8$ and $\mu_0 =2.8$, showing that the
  magnetic runs tend to form fewer but more massive objects (From
\citealt{VAZ05c}). }
\end{figure}

The influence of magnetic fields in controlling the SFE is a central
question in the turbulent model, given their fundamental role in the
AD-mediated model.  Numerical studies using ideal MHD,
neglecting AD, have shown that the field can only prevent
gravitational collapse if it provides {\em magnetostatic} support,
while MHD waves can only delay the collapse, but not prevent it 
\citep[][]{OST99,HEI01, LI04}.  
\cite{VAZ05c}
found a trend of decreasing SFE with increasing mean field strength in
\emph{driven} simulations ranging from non-magnetic to moderately
magnetically supercritical.  In addition, a trend to form fewer but
more massive objects in more strongly magnetized cases was also
observed (see Fig. \ref{fig:magnetic}).  Whole-cloud efficiencies
$\lesssim 5$\% were obtained for moderately supercritical regimes at
Mach numbers ${\cal M} \sim 10$. For comparison, \citet{NAK05} found
that comparable efficiencies in decaying simulations at
\emph{effective} Mach numbers ${\cal M} \sim $ 2--3 required
moderately subcritical clouds in the presence of AD. These results
suggest that whether MC turbulence is driven or decaying is a crucial
question for quantifying the role of the magnetic field in limiting
the SFE in turbulent MCs.

The relationship between the global and local SFE in galaxies still
needs to be elucidated.  The existence of the Kennicutt-Schmidt law
shows that global SFEs vary as a function of galactic properties, but
there is little indication that the local SFE in individual MCs varies
strongly from galaxy to galaxy.  Are they in fact independent?

\subsection{\textbf{Distribution of Clump Masses and Relation to the
 IMF}}
\label{sub:IMF}


The stellar IMF is a fundamental diagnostic of the star formation
process, and understanding its physical origin is one of the main
goals of any star formation theory.  Recently, the possibility that
the IMF is a direct consequence of the protostellar core mass
distribution (CMD) has gained considerable attention.
Observers of dense cores report 
that the slope at the high-mass end of the CMD resembles the
\citet{SAL55} slope of the IMF \citep[e.g.,][]{MOT98, TES98}. This has
been interpreted as evidence that those cores are the direct
progenitors of single stars.

\citet{PAD02} computed
the mass spectrum of self-gravitating density fluctuations
produced by isothermal 
supersonic turbulence with a log-normal density PDF and a power-law turbulent energy
spectrum with slope $n$, and proposed to identify this with the
IMF.  In this theory, the predicted slope of the IMF is given by
$-3/(4-n)$. For $n=1.74$, this gives $-1.33$, in agreement with the
Salpeter IMF, although this value of $n$ may not be universal
(see Section~\ref{sub:spectra_scal}). However, this theory suffers
from a problem common to any approach that identifies the core mass
function with the IMF: it implicitly assumes that the final stellar
mass is proportional to the core mass, at least on average. Instead,
high-mass cores have a larger probability to build up multiple stellar
systems than low-mass ones. 



Moreover, the number of physical processes that may play an important
role during cloud fragmentation and protostellar collapse is large.
In particular, simulations show that {\em (a)}~The mass distribution
of cores changes with time as cores merge with each other
\citep[e.g.,][]{KLE01c}. {\em (b)}~Cores generally produce not a
single star but several with the number stochastically dependent on
the global parameters \citep[e.g.,][]{KLE98, BAT05,DOB05}, even for low
levels of turbulence \citep{GOO04}.  {\em (c)}~ The shape of the clump
mass spectrum appears to depend on parameters of the turbulent flow,
such as the scale of energy injection \citep{SCH04} and the rms
Mach number of the flow \citep{BAL06}, perhaps because the density
power spectrum of isothermal, turbulent flows becomes shallower when
the Mach number increases \citep{CHO03, BER05, KIM05}. Convergence of
this result as numerical resolution increases sufficiently to resolve
the turbulent inertial range needs to be confirmed. A physical
explanation may be that stronger shocks produce denser, and thus
thinner, density enhancements. Finally, there may be other processes
like {\em (d)}~competitive accretion influencing the mass-growth
history of individual stars (see, e.g., \citealt{BAT05}, but for an
opposing view \citealt{KRU05c}), {\em (e)}~stellar feedback through
winds and outflows \citep[e.g.,][]{SHU87}, or {\em (f)}~changes in the
equation of state introducing preferred mass scales
\citep[e.g.,][]{SCA98b, LI03, JAP05, LAR05}.
Furthermore, 
even though some observational and theoretical studies of
dense, compact cores fit power-laws to the high-mass tail of the CMD,
the actual shape of those CMDs is often not a single-slope power law,
but rather a function with a continuously-varying slope, frequently
similar to a log-normal distribution 
(\citealt{BAL06}; see also Fig. 6 in the chapter by {\em Bonnell et
al.}).  So, the dynamic
range in which a power-law with slope $-1.3$ can be fitted is often
smaller than one order of magnitude.  All these facts call the
existence of a simple one-to-one mapping between the purely turbulent
clump-mass spectrum and the observed IMF into question.   

%
%
Finally, it is important to mention that, although it is generally
accepted that the IMF has a slope of $-1.3$ at the high-mass range,
the universal character of the IMF is still a debated issue, with strong
arguments being given both in favor of a universal IMF
\citep{ELM00c, KRO01b, KRO02, CHA05} and against it
in the Arches cluster \citep{STO05}, the Galactic Center region
\citep{NAY05} and near the central black hole of M31 \citep{BEN05}.

We conclude that both the uniqueness of the IMF and the relationship
between the IMF and the CMD are presently uncertain.  The definition
of core boundaries 
usually involves taking a density threshold. 
For a sufficiently high threshold, the CMD must reflect the IMF, while
for lower thresholds subfragmentation is bound to occur. Thus, future
studies may find it more appropriate to ask \emph{at what density
threshold} does a one-to-one relation between the IMF and the CMD
finally occur.

\subsection{\textbf{Angular Momentum and Multiplicity}}
\label{sub:ang-mom}

The specific angular momentum and the magnetic field of a collapsing
core determines the multiplicity of the stellar system that it
forms. We first review how core angular momentum is determined, and
then how it in turn influences multiplicity.

\vskip 0.3cm
\noindent{{\em 5.3.1. Core Angular Momentum.}}
Galactic differential rotation corresponds to a specific angular momentum $j
\approx  10^{23}\,$cm$^2$\,s$^{-1}$, while on scales of cloud cores,
below $0.1\,$pc,  $j \approx 10^{21}\,$cm$^2$\,s$^{-1}$. A
binary G star with an orbital period of 3 days has $j \approx
10^{19}\,$cm$^2$\,s$^{-1}$, while the spin of a typical T Tauri star
is a few $\times 10^{17}\,$cm$^2$\,s$^{-1}$. Our own Sun rotates only
with $j\approx 10^{15}\,$cm$^2$\,s$^{-1}$.  Angular momentum must be
lost at all stages \citep[for a review, see][]{BOD95}.

Specific angular momentum has been measured in clumps and cores at
various densities with different tracers:
{\em (a)} $n \sim 10^3 \mbox{ cm}^{-3}$ observed in $^{13}$CO
\citep{ARQ86}; 
{\em (b)}  $n \sim 10^4 \mbox{ cm}^{-3}$ observed in NH$_3$ (e.g.,
\citealt{GOO93, BAR98}; see also \citealt{JIJ99}); 
{\em (c)} $n \sim 10^4-10^5 \mbox{ cm}^{-3}$ observed in N$_2$H$^+$ in
both low \citep{CAS02} and high \citep{PIR03} mass star forming
regions.
In all these objects, the rotational energy is
considerably lower than required for support; in the densest cores the
differences is an order of magnitude. \cite{JAP04} showed that these
clumps form an evolutionary sequence with $j$ declining with
decreasing scale.  Main sequence binaries have $j$ just below that of
the densest cores.

Magnetic braking has long been thought to be the primary mechanism for
the redistribution of angular momentum out of collapsing objects
\citep{EBE60,MOU79,MOU80}. Recent results suggest that magnetic
braking may be less important than was thought.  Two groups have now
performed MHD models of self-gravitating, decaying \citep{GAM03} and
driven \citep{LI04} turbulence, finding
distributions of specific angular momentum consistent with observed
cores and with each other.  Surprisingly, however, \citet{JAP04} found
similar results with {\em hydrodynamic} models \citep[see
also][]{TIL04}, suggesting that 
accretion and gravitational interactions may dominate over magnetic
braking in determining how angular momentum is lost during the
collapse of cores. 
\citet{FIS04a} reached similar conclusions based
on semi-analytical models.

\vskip 0.3cm
\noindent{{\em 5.3.2. Multiplicity.}}
Young stars frequently have a higher multiplicity than main sequence
stars in the solar neighborhood  \citep{DUC99, MAT00}, suggesting that
stars are born with high multiplicity, but 
binaries are then destroyed dynamically during their early lifetime
\citep{SIM93, KRO95b, PAT98}. 
In addition, multiplicity depends on mass.  Main sequence stars with
lower masses have lower multiplicity \citep{LAD06}.

Fragmentation during collapse is considered likely to be the dominant
mode of binary formation \citep{BOD00}.  Hydrodynamic models showed
that isothermal spheres with a ratio of thermal to gravitational
energy $\alpha < 0.3$ fragment \citep{MIY84, BOS93, BOS95, TSU99,
TSU99b}, and that even clumps with 
larger $\alpha$ can still fragment if they rotate fast enough to form
a disk \citep{MAT03}. 

The presence of protostellar jets
driven by magnetized accretion disks \citep[e.g.,][]{KON93} 
strongly suggests that magnetic
braking must play a significant role in the final stages of collapse
when multiplicity is determined.  Low resolution MHD models showed
strong braking and no fragmentation \citep{DOR82, BEN84b, PHI86,
PHI86b, HOS04}.  \citet{BOS00, BOS01, BOS02, BOS05}
found that fields enhanced fragmentation, but neglected magnetic
braking by
only treating the radial component of the magnetic tension force. \citet{ZIG05}
showed that field strengths small enough to allow for binary formation
cannot provide support against collapse, offering additional support
for a dynamic picture of star formation.

High resolution models using up to 17 levels of nested grids have now
been done for a large number of cores \cite{
TOM05, 
MAC05b}.  These clouds were initially cylindrical, in hydrostatic
equilibrium, threaded by a magnetic field aligned with the rotation
axis having uniform plasma $\beta_p$ (the ratio of magnetic field to
thermal pressure; \citealt{MAC04b}).
At least for these initial
conditions, fragmentation and binary formation happens during collapse
if the initial ratio
\begin{equation}
\frac{\Omega_0}{B_0} > \frac{G^{1/2}}{2^{1/2} c_s} \sim 3 \times 10^7
\mbox{ yr}^{-1}\mu\mbox{ G}^{-1}.
\end{equation}
Whether this important result generalizes to other geometries 
clearly needs to be confirmed.

If prestellar cores form by collapse and fragmentation in a turbulent
flow, then nearby protostars might be expected to have aligned angular
momentum vectors as seen in models \citep{JAP04}.  This appears to be
observed in binaries \citep{MON06}, and stars with ages less
than $10^6$~yr. 
However, the effect
fades with later ages \citep[e.g.,][]{MEN04}.   \citet{MEN04} also make the
intriguing suggestion that jets may only appear if disks are aligned
with the local magnetic field.


\subsection{\textbf{Importance of Dynamical Interactions During
     Star Formation}}
\label{sub:dynamics}


Rich compact clusters can have large numbers of protostellar objects in
small volumes.  Thus, dynamical interactions between protostars may
become important, introducing a further degree of stochasticity to the
star formation process besides the statistical chaos associated with
the gravoturbulent fragmentation process.


Numerical simulations have shown that when a MC region of a few
hundred solar masses or more coherently becomes gravitationally
unstable, it contracts and builds up a dense clump highly structured
in a hierarchical way, containing compact protostellar
cores \citep[e.g.,][]{KLE00a,KLE01b,BAT03,CLA05}.  Those cores may contain
multiple collapsed objects that will compete with each other for
further mass accretion in a common limited and rapidly changing
reservoir of contracting gas. 
The relative importance of these competitive processes depends on the
initial ratio between gravitational vs.\ turbulent and magnetic
energies of the cluster-forming core (\citealt{KRU05c}; see also the chapter by
 {\em Bonnell et al.}). If gravitational contraction strongly dominates the
dynamical evolution, as 
seems probable for 
the formation of massive,
bound clusters, then the following effects need to be considered.

%

{\em (a)}~Dynamical interactions between collapsed objects during the
embedded phase of a nascent stellar cluster evolve towards energy
equipartition.  As a consequence, massive objects will have, on
average, smaller velocities than low-mass objects.  They sink towards
the cluster center, while low-mass stars predominantly populate large
cluster radii. This effect already holds for the embedded phase
\citep[e.g.,][]{BON98a,KLE00a,BON04}, and agrees with observations of
young clusters \citep[e.g., ][ for NGC330 in the Small Magellanic
Cloud]{SIR02}.


{\em (b)}~The most massive cores within a large clump have the largest
density contrast and collapse first.  In there, the more massive
protostars begin to form first and continue to accrete at a high rate
throughout the entire cluster  evolution ($\sim 10^5$\,yr).
%
%
As their parental cores merge with others, more gas is fed into their
`sphere of influence'. They are able to maintain or even increase the
accretion rate when competing with lower-mass objects
\citep[e.g.,][]{SCH04,BON04}.
%
%

%


{\em (c)}~The previous processes lead to highly time-variable
protostellar mass growth rates \citep[e.g.,][]{BON01a,KLE01a,SCH04}.
As a consequence, the resulting stellar mass spectrum can be modified
(\citealt{
FIE65, SIL79,LEJ86,MUR96,BON01b,KLE01c,SCH04,BAT05}; see also the
chapters by {\em Bonnell et al.}\ and {\em Whitworth et al.}).
\citet{KRU05c} argue that these works typically do not resolve the
Bondi-Hoyle radius well enough to derive quantitatively correct
accretion rates, but the qualitative statement appears likely to
remain correct.
%


{\em (d)}~Stellar systems with more than two members are in general
unstable, with the lowest-mass member having
the highest probability of being expelled \citep[e.g.,][]{vALB68}.

%

{\em (e)}~Although stellar collisions have been proposed as a
mechanism for the formation of high-mass stars \citep{BON98b,STA00},
detailed 2D and 3D calculations \citep[][]{YOR02,KRU05} show that mass
can be accreted from a protostellar disk onto the star.  Thus,
collisional processes need not be invoked for the formation of
high-mass stars \cite[see also,][]{KRU05c}.

%

{\em (f)}~Close encounters in nascent star clusters can truncate
and/or disrupt the accretion disk expected to surround (and feed) every
protostar.  This influences mass
accretion through the disk, modifies the ability to subfragment and
form a binary star, and the probability of planet formation
\citep[e.g.,][]{CLA91,McD95,SCALLY01,KRO01c,BON01d}.  In particular,
\citet{IDA01} note that an early stellar encounter may explain
features of our own solar system, namely the high eccentricities and
inclinations observed in the outer part of the Edgeworth-Kuiper Belt
at distances larger than $42\,$AU.


{\em (g)} While competitive coagulation and accretion is a viable
mechanism in very massive and dense star-forming regions,
protostellar cores in low-mass, low-density clouds such as Taurus or
$\rho$-Ophiuchus are less likely to strongly interact with each other.

 \subsection{\textbf{Effects of Ionization and Winds}}
 \label{sec:destruction}

When gravitational collapse proceeds to star formation, ionizing
radiation begins to act on the surrounding MC.  This can drive
compressive motions that accelerate collapse in surrounding gas, or
raise the energy of the gas sufficiently to dissociate molecular gas
and even drive it out of the potential well of the cloud, ultimately
destroying it.  Protostellar outflows can have similar effects, but
are less powerful, and probably do not dominate cloud energy
(\citealt{MAT02,NAK05}; see also the chapter by {\em Arce et al.}\
2006). 

The idea that the expansion of H~{\sc ii} regions can compress the
surrounding gas sufficiently to trigger star formation was proposed by
\citet{ELM77}.  Observational and theoretical work supporting it was
reviewed by \cite{ELM98}.  Since then, it has become reasonably clear that
star formation is more likely to occur close to already formed stars
than elsewhere in a dark cloud \citep[e.g.,][]{OGU02, STA02, KAR03,
BAR03, CLA04b, HEA04, DEH05}.  However, the question remains whether
this seemingly triggered star formation would have occurred anyway in
the absence of the trigger.  Massive star formation occurs in
gravitationally collapsing regions where low mass star formation is
anyway abundant.  The triggering shocks produced by the first massive
stars may determine in detail the configuration of newly formed stars
without necessarily changing the overall star formation efficiency of
the region \citep{ElM02b, HES04}. 
Furthermore, the energy input by the shock fronts has the net effect
of inhibiting collapse by driving turbulence \citep{MAT02, VAZ03, MAC04},
even if it can not prevent local collapse (see Section~\ref{sec:SFE}).

The study of the dissociation of clouds by ionizing radiation also has
a long history, stretching back to the 1D analytic models of
\citet{WHI79} and \citet{BOD79b}. The latter group first noted the
champagne effect, in which ionized gas in a density gradient can blow
out toward vacuum at supersonic velocities.  This mechanism was also
invoked by \citet{BLI80}, and further studied by \citet{FRA94} and
\citet{WIL97}. \citet{MAT02} has performed a detailed analytic study
recently, which suggests that the energy injected by H~{\sc ii}
regions is sufficient to support GMCs for as long as $2 \times
10^7$~yr.  One major uncertainty remaining is whether expanding H~{\sc
ii} regions can couple to the clumpy, turbulent gas as well as is
assumed.


\section{\textbf{CONCLUSIONS}}\label{sec:conclusions}

We have reviewed recent work on MC turbulence and
discussed its implications for our understanding of star formation.
%
There are several aspects of MC and star formation where our
understanding of the underlying physical processes has developed
considerably, but there are also a fair number of questions that 
remain unanswered. The following gives a brief overview.

{\em (a)}~MCs appear to be transient objects, representing the
brief molecular phase during the evolution of dense regions formed by
compressions in the diffuse gas, rather than long-lived, equilibrium
structures.

{\em (b)}~ Several competing mechanisms for MC formation remain
viable.   
At the largest scales, the gravitational instability of the combined
system of stars plus gas drives the formation of spiral density waves
in which GMCs can form either by direct gravitational instability of
the diffuse gas (in gas-rich systems), or by compression in the
stellar potential followed by cooling (in gas poor systems). At
smaller scales, supersonic flows driven by the ensemble of SN
explosions have a similar effect.

{\em (c)}~
Similar times ($< 10-15$~Myr) are needed both to assemble enough gas
for a MC and to form molecules in it.  This is less than earlier
estimates.  Turbulence-triggered thermal instability can further
produce denser ($10^3$~cm$^{-3}$), compact H~I clouds, where the
production of molecular gas takes $\lesssim 1$~Myr.  Turbulent 
flow through the dense regions results in broad regions of molecular
gas at lower average densities of $\sim$~100~cm$^{-3}$.

{\em (d)}~Interstellar turbulence seems to be driven from 
scales substantially larger than MCs. Internal MC turbulence may
well be a byproduct of the cloud formation mechanism itself, explaining
why turbulence is ubiquitously observed on all scales.

{\em (e)}~MC cloud turbulence appears to produce a well-defined
velocity-dispersion to size relation at the level of entire MC
complexes, with an index $\sim 0.5$-0.6, but becoming less well
defined as smaller objects are considered. The apparent density-size
relation, on the other hand, is probably an observational artifact.

{\em (f)}~MC cores are produced by turbulent
compression and may, or may not, undergo gravitational instability.
They evolve dynamically on timescales of $\sim 10^6$ yr, but still can
mimic certain observational properties of quasistatic objects, such as
sub- or transonic non-thermal central linewidth, or Bonnor-Ebert
density profiles.  These results agree with cloud statistics of
nearby, well-studied regions.
%

{\em (g)}~
Magnetic fields appear unlikely to be of qualitative importance
in structuring MC cores, although they probably do make a quantitative
difference. 

{\em (h)}~
Comparison of magnetized and unmagnetized simulations of
self-gravitating, turbulent gas reveals similar angular momentum
distributions.  The angular momentum distribution of MC cores and
protostars may be determined by turbulent accretion rather than
magnetic braking.

{\em (i)}~The interplay between supersonic turbulence and gravity on
Galactic scales can explain the global efficiency of converting warm
atomic gas into cold molecular clouds. The local efficiency of
conversion of MCs into stars, on the other hand, seems to depend
primarily on stellar feedback
evaporating and dispersing 
the cloud, and only secondarily on the presence of magnetic fields.

{\em (j)}~There are multiple reasons to doubt the existence of a
simple one-to-one mapping between the purely turbulent clump mass
spectrum and the observed IMF, despite observations showing that the
dense self-gravitating core mass spectrum may resemble the IMF. A
comprehensive understanding of the physical origin of the IMF remains
elusive.

{\em (k)}~In dense clusters, star formation 
might 
be affected by
dynamical interactions.  Close encounters between newborn stars and
protostellar cores 
have the potential to 
modify their accretion history, and consequently
the final stellar mass, as well as the size of protostellar disks, and
possibly their ability to form planets and planetary systems.


\smallskip

 \textbf{ Acknowledgments.} We acknowledge useful comments and suggestions 
 from our referee, Paolo Padoan.  JB-P and EV-S were supported by
 CONACYT grant 36571-E.  RSK was supported by the Emmy Noether Program
 of the DFG under grant KL1358/1.  M-MML was supported by NSF grants
 AST99-85392, AST03-07793, and AST03-07854, and by NASA grant NAG5-10103.
 
{\small


\end{document}